# Improvement of Identification Procedure Using Hybrid Cuckoo Search Algorithm for Turbine-Governor and Excitation System

Teimour Hosseinalizadeh, S. Mahmoud Salamati, S. Ali Salamati, and G. B. Gharehpetian

*Abstract*— In this paper a new method is introduced in order to modify identification process of a gas power plant using a meta-heuristic algorithm named Cuckoo Search (CS). Simulations play a significant role in dynamic analyses of power plants. This paper points out to a practical approach in model selection and parameter estimation of gas power plants. The identification and validation process concentrates on two subsystems: governor-turbine and exciter. Standard models GGOV1 and STB6 are preferred for the dynamical structures of governor-turbine and exciter respectively. Considering definite standard structure, main parameters of dynamical model are pre-estimated via system identification methods based on field data. Then obtained parameters are tuned carefully using an iterative Cuckoo algorithm. Models must be validated by results derived via a trial and error series of simulation in comparison to measured test data. The procedure gradually yields in a valid model with precise estimated parameters. Simulation results show accuracy of identified models. Besides, a whiteness analysis has been performed in order to show the authenticity of the proposed method in another way. Despite various detailed models, practical attempts of model selection, identification, and validation in a real gas unit could rarely be found among literature. In this paper, Chabahar power plant in Iran, with total install capacity of 320 MW, is chosen as a benchmark for model validation.

## NOMENCLATURE

| | |
|---|---|
| $R$ | Permanent droop |
| $T_{Pelec}$ | Electrical power transducer time constant |
| $K_{pgov}$ | Governor proportional gain |
| $K_{igov}$ | Governor integral gain |
| $K_{dgov}$ | Governor derivative gain |
| $T_{dgov}$ | Governor derivative controller time constant |
| $T_{act}$ | Actuator time constant |
| $K_{turb}$ | Turbine gain |
| $W_{fnl}$ | No load fuel flow |
| $T_b$ | Turbine lag time constant |
| $T_c$ | Turbine lead time constant |
| $T_{eng}$ | Transport time delay for diesel engine |
| $T_{fload}$ | Load limiter time constant |
| $K_{pload}$ | Load limiter proportional gain for PI controller |
| $K_{iload}$ | Load limiter integral gain for PI controller |
| $L_{dref}$ | Load limiter reference value |
| $Dm$ | Speed sensitivity coefficient |
| $K_{imw}$ | Power controller gain |
| $P_{mwset}$ | Power controller set-point |
| $K_a$ | Acceleration limiter gain |
| $T_a$ | Acceleration limiter time constant |
| $I_{FD}$ | Synchronous machine field current |
| HV Gate | Model block with two inputs and one output, the output always corresponding to the higher of the two inputs |
| LV Gate | Model block with two inputs and one output, the output always corresponding to the lower of the two inputs |
| $E_{FD}$ | Exciter output voltage |
| $I_{LR}$ | Exciter output current limit reference |
| $V_C$ | Output of terminal voltage transducer and load compensation elements. |
| $V_{REF}$ | Voltage regulator reference voltage |
| $V_B$ | Available exciter voltage |
| $K_{CI}$ | Exciter output current limit adjustment |
| $K_M$ | Forward gain constant of the inner loop field regulator. |
| $K_{PA}, K_{IA}$ | Voltage regulator proportional and integral gains |
| $K_{FF}$ | Pre-control gain constant |
| $K_{LR}$ | Exciter output current limiter gain |

## I. INTRODUCTION

DYNAMIC performance analyses of power system have been planned by engineers since 1970's [1]. Gas power plants consist of many complicated subsystems dealing with high installation and maintenance expenditures. Practical evaluation, experimental labor, and engineering costs in the improvement of such systems are expected to invest large amount of money. Therefore, subsystem developments sometimes are found impossible due to economic reasons. On the other hand, operational limitations may yield in technological infeasibility of new updates in control structures. Hence, a simulator tool capable of realistic approximation and transient performance analyses is of researchers' interest.

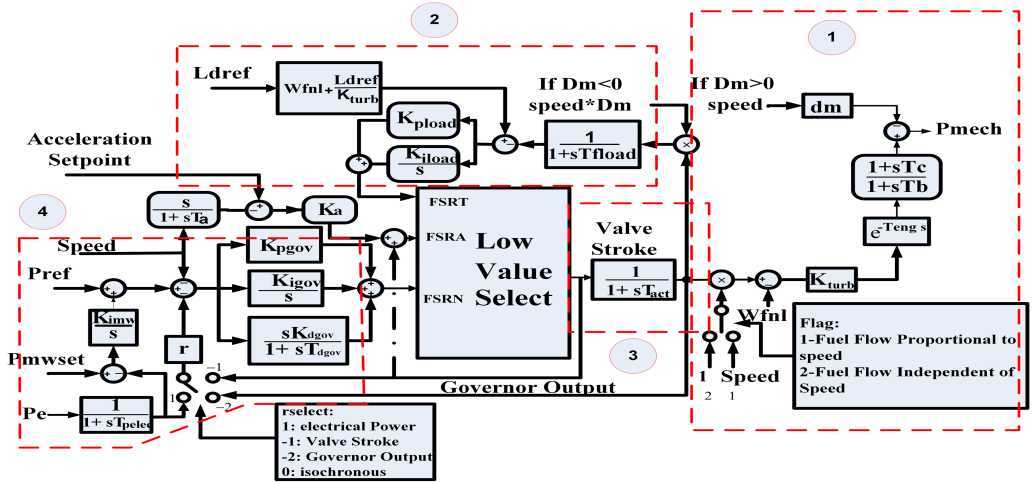

**Fig. 1. GGOV1 IEEE standard model.**

Dynamic models of different parts in a gas unit comprise of differential and algebraic equations derived according to basic principles governing system phenomenon.

Various approaches with a wide range of complexity are available in this field. Some of these approaches present a complete model of total behavior of power plant [2]-[7]. A deductive approach is used in [2] for turbine modeling as a part of integrated power system model for steady state performance verification. In [3], a multi-temporal simulation model is implemented in order to carry out integrated analysis of electricity, heat, and gas distribution. The main disadvantage of the model proposed in [3], is the complexity of relevant coupled electrical, heat and gas flow equations requiring to be solved simultaneously via Newton-Raphson approach. In [4], similar attempts as [3] are performed for biomass power plant using micro gas turbine. In [7], a power generating unit dynamic model is proposed in order to investigate boiler pressure effects, load-frequency control, boiler-turbine and coordinated control tuning. Different subsystems are modeled and identified individually via second order transfer function approach, not considering any special standard in model presentation. Partial modeling of a special subsystem like governor-turbine [8], exciter [9], and generator [10] are also found in the literature. By the way, modeling of integrated thermal power plants or their individual equipment could be categorized in two groups; white-box and black-box models. White-box models deal with dynamic equations [11], while black-box models are employed when access to dynamic equations is impossible [12]. Due to disadvantages like over parameterization, training complexity, and lack of practicality, we utilize a white-box scheme in order to model the subsystems of the thermal power plant. In fact, IEEE standard models are used in this paper and their parameters are identified using identification techniques. Models for two main parts of a gas power plant including governor-turbine and exciter are identified and validated in this work.

Thermal governor modeling approaches are common research topic in the literature. However, practical model validation of presented schemes has been rarely performed. Western Electricity Coordinating Council (WECC) has started an important research on validation of governor-turbine dynamic models. A new modeling approach is proposed and has been extensively validated against recording from WECC benchmark systems. It is concluded based on results that presented approach could be implemented in un-responsive characteristics of frequency control simulation via past models. Therefore, this model has been approved for use in all operation and planning studies in WECC.

There exist different excitation system models considering detailed system properties in the literature from direct current (DC) to static (ST) types; however, in a practical model derivation and validation, facility in obtaining model parameters from field test data is very important.

The paper is organized as follows: governor-turbine and exciter standard models are discussed in Section II and III, respectively. In section IV an easily powerful metaheuristic algorithm is described. System setup and proposed identification procedure using field data are discussed in section V. Also model training results and validations are given in this section. At last, section VI concludes the paper and recommends some practical remarks in order to improve the GGOV1 model.

## II. GOVERNOR-TURBINE MODELING

Several documents published in the last decade give various modeling approaches for dynamics behavior of governor-turbine. A hierarchy of models are presented for heavy-duty gas turbines in [13]. Models categorized in this standard include GAST, the most simplistic representation of a gas-turbine, GAST2A, considering a proportional speed governor control, GGOV1, developed as a general purpose of dynamic simulation studies.

Fig. 1 demonstrates the IEEE standard GGOV1 governor-turbine model which is our chosen model for parameter identification. GGOV1 is flexible to model various governor model and feedback signal. This is the main advantage of GGOV1 in comparison to GAST and GAST2A. This model is suitable for grid operation studies in large networks and load rejection analyses that were impractical in previous models.

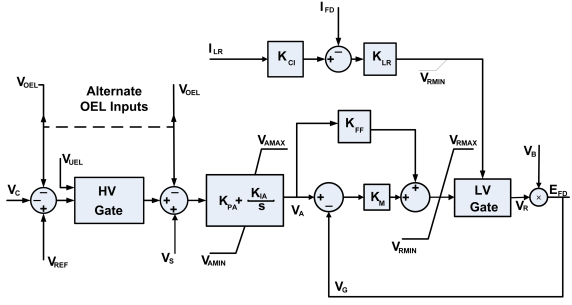

Fig. 2. ST6B IEEE Exciter standard model.

In this standard model, electrical power, governor output, or valve stroke can be utilized as feedback droop signal. The input of valve model is derived from a Low-Select block which selects between "FSRN", "FSRT", and "FSRA" signals denoting speed governor, temperature, and acceleration control signal respectively. GGOV1 represents all fundamental elements of a gas-turbine controller. Speed/power control normally is performed between about 70-100 percent of nominal load. The other two control loops are active in maximum load enforcement. Detailed structure of model is explained in [13]. Acceleration controller loop is not implemented in some gas power plants such as Chabahar power plant; so there is no discussion about its parameters in this paper. Also droop control loop gets electrical power ($p_e$) to adjust turbine speed. GGOV1 model is partitioned into four different subsystems which are determined in Fig. 1. So, obtaining of these parts' parameters are necessary in order to evaluate the gas turbine behavior.

## III. EXCITATION SYSTEM MODELING

Various excitation models could be found in the previous literature [14], [15]. However, excitation system models suitable for use in large-scale system stability studies must be selected for practical attempt of power plant model validation.

In 421.5-2005 Standard of IEEE recommended models for practical stability studies are listed. The model structures presented are intended to facilitate the use of field test data as a means of obtaining model parameters. In 421.5-2005 Standard, exciter models are presented in three categories [16]: AC, DC, and Static. Exciter system in Chabahar power plant is of static type. Hence, we must choose an appropriate model for experimental study of static category. The model selected for excitation system in this paper is ST6B. The structure of ST6B is illustrated in Fig. 2. Low-value Gate block in the model has two inputs which one of them is constructed by $I_{FD}$ and $I_{LR}$ is not active during normal condition. So the main focus in this paper is identifying of the exciter model in normal operation. In some power plant this model is called AVR mode.

## IV. CUCKOO SEARCH ALGORITHM

Metaheuristic algorithms such as Genetic and Particles Swarm Optimization (PSO) are nature based algorithms which have been used in wide range of problems [17]. Cuckoo Search algorithm is one of these metaheuristic methods which has been shown to be effective in solving even non-polynomial time (NP) problems. One of the main advantages of this algorithm is that there are fewer parameters needed to be tuned in CS than in PSO and GA. Furthermore, it has been shown for multimodal objective function CS has better performance than GA [18], [19]. Although CS has other details but pseudo code for its main form is as follows

CS algorithm via Lévy flights [20]
**Begin**
  Objective function $f(x), x = (x_1, \ldots, x_d)^T$
  Produce initial population of $n$ host nests $x_i$ $i = (1, \ldots, n)$
  **While**($t$<MaxGeneration) or (stop criterion)
  Get a Cuckoo randomly by Lévy flights
  Evaluate it fitness $F_i$
  Choose a nest among $n$ (say $j$) randomly
   **If** ($F_i > F_j$)
   Replace $j$ by the new solution
   **end**
  A fraction ($p_a$) of worst nests are abandoned and new ones are built.
  Keep the best solutions
  Rank the solutions and find the current best
  **end while**
**end**

Generating new solution $x_i^{(t+1)}$ is performed using Lévy flight which provides a random walk with below equation

$$x_i^{(t+1)} = x_i^{(t)} + a \oplus Lévy(\lambda) \qquad (1)$$

Where $\alpha$ is the step size, which is set at $\alpha = 1$ in this paper, the product $\oplus$ means entrywise multiplication and $Lévy$ is a distribution which has infinite mean and variance and its distribution is

$$Lévy \sim u = t^{-\lambda}, \qquad (1 < \lambda \le 3). \qquad (2)$$

In CS algorithm $p_a$ is the probability that the egg laid by a cuckoo is discovered by the host bird and we set this parameter in this paper at 0.25. Maximum generation and stop criterion are set at 100 and $e^{-2}$ for all simulation in the next section. Also appropriate initial population will be useful in reducing algorithm convergence time. This algorithm is used in the proposed identification procedure in the next section.

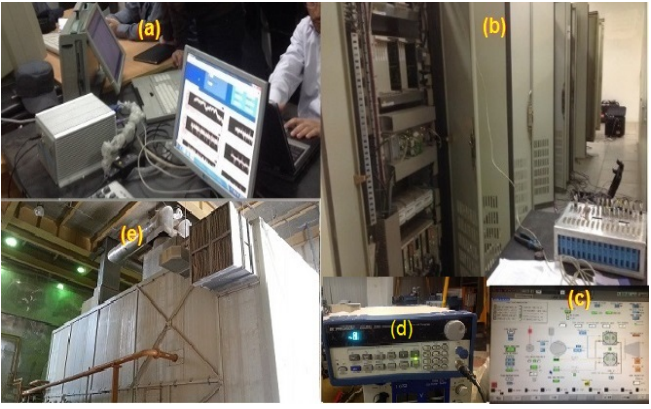

Fig. 3. Test setup in Chabahar gas power plant. (a) high speed data acquisition system (b) PLC setup (c) HMI of gas turbine and governor system (d) analog signal generator used for excitation in test time (e) generator room.

## V. SYSTEM IDENTIFICATION

### A. System setup

To identify and validate parameters in GGOV1 and ST6B standard models different tests have been performed on the gas power plant while it was connected to the main grid. Excitation signal has a great influence in quality of identified model parameters. Although it has been recommended to use signal with rich frequency content (known as persistency of excitation) such as pseudo random binary signal (PRBS) and white noise [21], these signals are not applicable in identification of power plant subsystems because their sudden changes in value and frequency raise practical issue about system protection. So instead of using PRBS, square pulse wave with almost 5MW changes in power set point was applied to the power reference and all input and output of the subsystem has been recorded by 1 ms sample rate. Furthermore generator and gas turbine main specification are listed in Appendix.

Fig. 3 Shows different part of the Chabahar gas power plant test setup. Data acquisition system which is used for recording data with high sample rate is shown in part (a) of this figure. Part (b) is power plant central controller, part (c) is human machine interface (HMI) which is used for monitoring of power plant.

Analog signal generator used for system excitation is shown at (d) and generator with gas turbine overall room is displayed at (e). Using above setup two different data sets were collected for parameter estimation by proposed algorithm and validation of obtained parameters.

### B. Identification method

The proposed identification procedure is explained in the following steps.

1. Data collection. In this part, input and output data for each subsystem has to be gathered using appropriate sample rate. All test are performed with square pulse signal generated by industrial signal generator system.
2. Collected data should be processed. In most cases there is environmental noise which has to be removed from the actual signal and also data has to be per unitized by their ranges to have per-unitized parameters. A low-pass Butterworth filter with cutoff frequency $40 Hz$ is used for filtering data.
3. In this stage least square method is used for pre-identifying desired parameters. In this approach parameters are estimated by

$$\theta = (X^T X)^{-1} X^T Y \quad (3)$$

Where $Y$ is $N \times 1$ output vector for each subsystem, $X$ is $N \times t$ regressor matrix and $\theta$ is $t \times 1$ vector of subsystem model parameters.

4. Use identified parameter in the previous step as initial population for CS algorithm. Error index which is used as objective function in CS is mean square error type.

$$error = \frac{1}{N} \sum_{i=1}^{N} (y_i - \hat{y}_i)^2 \quad (4)$$

Where $N$ is data total number, $y_i$ is $i$th sample of recorded output data and $\hat{y}_i$ is $i$th sample of simulated output data.

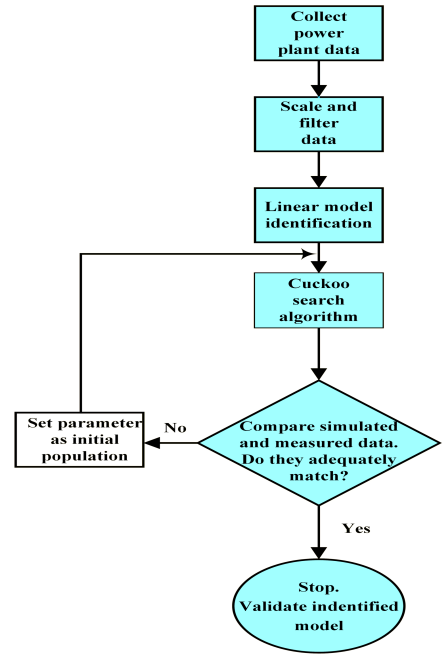

Fig. 4. Flowchart of proposed identification method

If the results is satisfying, end the algorithm and use different data set to validate identified parameters, otherwise apply CS results as initial population and go to step 4.

Fig. 4 demonstrates the used flowchart for parameter estimation in this paper. We should notice although algorithm can work without linear identification section but convergence time for parameter estimation will be longer in this way. So one can consider this algorithm as a hybrid optimization method in the context of metaheuristic approaches in solving similar problems.

C. *Mo*del training results

As mentioned in part A of this section field data recorded when the power plant was connected to the main grid. So for applying proposed identification algorithm there is no need to plant isolation from the main grid. Subsystem training results are shown in Fig.5. As one can notice proposed identification

TABLE I
ERROR INDEX VALUES FOR TRAINING DATA

| System part | Error index(percent) |
|---|---|
| Subsystem (1) | 0.0239 |
| Subsystem (2) | 0.0268 |
| Subsystem (3) | 0.001 |
| Subsystem (4) | 0.1192 |
| Exciter (5) | 0.0293 |

D. *Model validation results*

Fig. 6 indicates validation results of the identified parameters using different data sets. Although tests have been performed in different operating point of the gas turbine but Fig. 6 (a) and Fig 6 (b) show identified parameters have appropriate dynamical responses so that they are acceptable. Test results also show both GGOV1 and ST6B standard models have very good conformity with real gas turbine and exciter system.

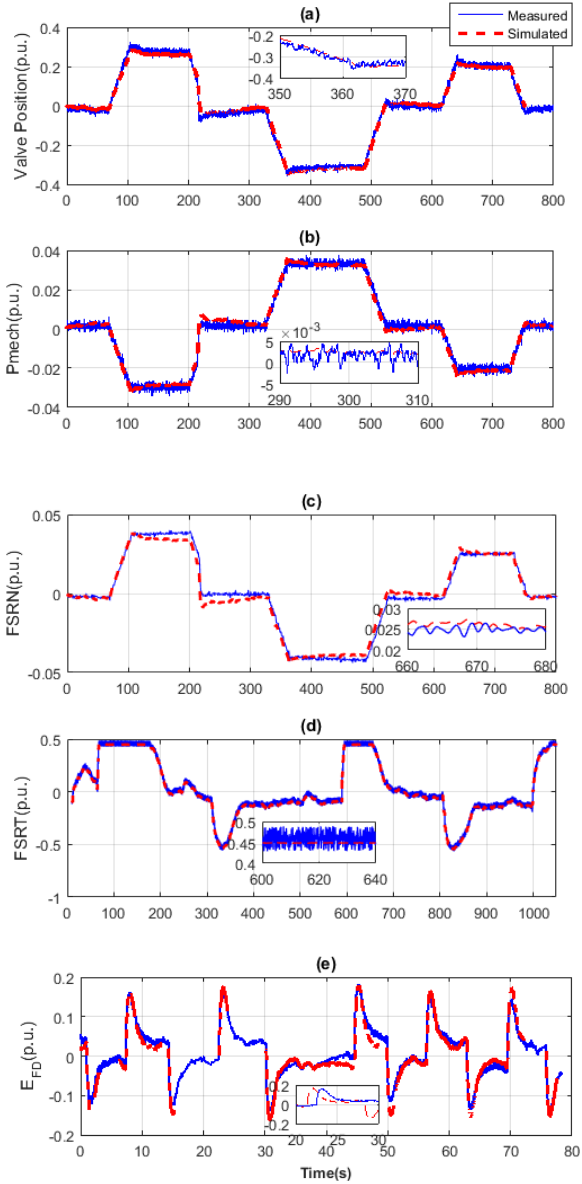

Fig. 5. Measured and simulated results for different part of the system using training data. (a) Valve position (b) electrical power (c) speed controller output (d) temperature controller output (e) exciter output.

algorithm successfully estimated unknown parameters and also results are displayed in per unit based so results can be compared easily. Table I also indicates the error index values in percent for different subsystems. All subsystems have error indices less than 0.2 percent which this result confirms conclusions have been inferred from Fig. 5.

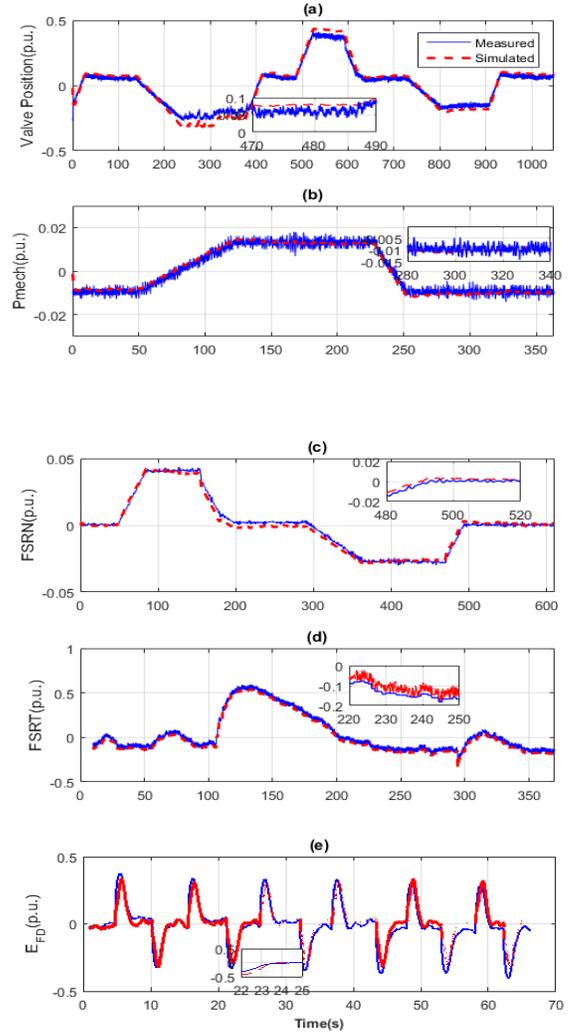

Fig. 6. Measured and simulated results for different part of the system using validation data. (a) Valve position (b) electrical power (c) speed controller output (d) temperature controller output (e) exciter output.

Error index for all subsystems are given in Table II. As expected error values for validation indices are higher than training counterparts but they are still below the 0.5 percent so we conclude identified models have not been experts just for training data. At the end, in our opinion, parameter identification for studied power plant was done successfully.

TABLE II
ERROR INDEX VALUES FOR VALIDATION DATA

| System part | Error index(percent) |
|---|---|
| Subsystem (1) | 0.1176 |
| Subsystem (2) | 0.0343 |
| Subsystem (3) | 0.0614 |
| Subsystem (4) | 0.1210 |
| Exciter (5) | 0.4085 |

*E. Whiteness test*

Although defined error indices indicate that identified models have good qualities, but they do not show anything about what couldn't be identified i.e. if there is any lost data in the residuals (errors) or not. Auto correlation for the residuals is as follows

$$\hat{R}_e^N(\tau) = \frac{1}{N}\sum_{t=1}^{N} e(t)e(t-\tau) \qquad (5)$$

Where $e(t) = y(t) - \hat{y}(t)$. If (5) is small for $\tau \neq 0$ then one can conclude residuals are white noise so models' quality can be proved in another way. To verify similarity between white noise and residuals using (5) one should check this inequality

$$\frac{N}{\left(\hat{R}_e^N(0)\right)^2} \sum_{\tau=1}^{M}\left(\hat{R}_e^N(\tau)\right)^2 < \beta^2 \ . \qquad (6)$$

$\beta$ is computed using $\frac{1}{\sqrt{2\pi}}\exp\left(\frac{-\beta^2}{2}\right) = \alpha$ where $\alpha$ is defined confidence level that is set $\alpha = 0.01$. Details for this method can be found in [21].

Fig. 7 and Fig. 8 show defined index and confidence levels for training and validation data, respectively. As we see residual autocorrelations for different models are located between confidence levels so (6) is confirmed and correctness of identified parameters are validated.

As an alternative of CS algorithm, GA and PSO have been used in proposed identification method to compare their performances with CS algorithm in our defined problem.

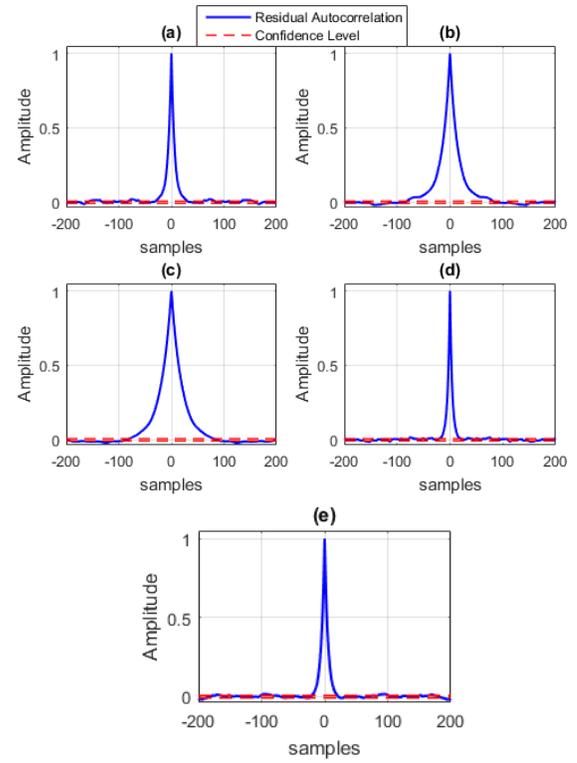

Fig. 7. Autocorrelation using training data for (a) valve position (b) electrical power (c) speed controller (d) temperature controller (e) exciter, identified models.

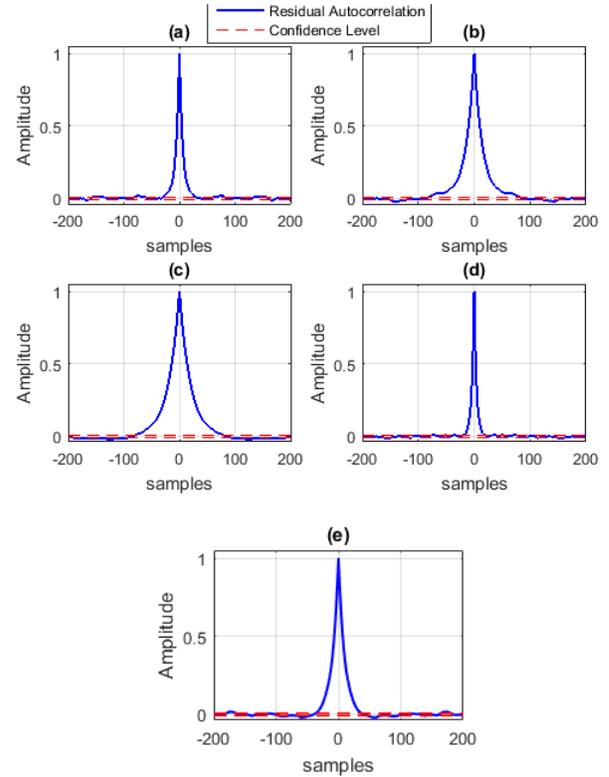

Fig. 8. Autocorrelation using validation data for (a) valve position (b) electrical power (c) speed controller (d) temperature controller (e) exciter, identified models.

GA and PSO have many parameters as degrees of freedom that must be tuned independently for each problem. The best obtained results according to error indices are shown in Fig. 9 and estimated parameters are included at Table III for different algorithms.

As it can be seen from Fig. 9, PSO algorithm has higher error indices in comparison to CS and GA, therefore, its estimated parameters are not acceptable in this problem. CS and GA results are approximately similar to each other. Actually in this problem there is not a major difference in CS and GA performances so one can use them for parameter identification problems. Although, as described before, due to its relative simplicity in tuning of required parameters, CS can be the first choice for solving similar problems.

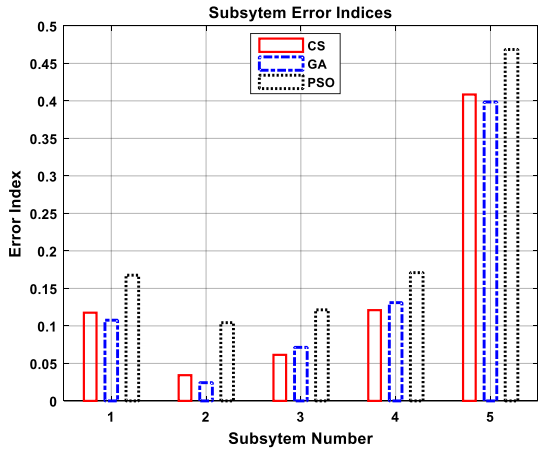

Fig. 9. Validation data error indices for CS, GA and PSO algorithms for defined system parts

TABLE III
IDENTIFIED SYSTEM PARAMETER VALUES

| System parameter | CS | GA | PSO |
|---|---|---|---|
| $K_{pgov}$ | 3.10 | 3.12 | 2.9 |
| $K_{igov}$ | 0.90 | 0.91 | 0.95 |
| $K_{dgov}$ | 0.0 | 0.0 | 0.10 |
| $T_{dgov}$ | 0.0 | 0.0 | 0.15 |
| $T_{act}$ | 1.83 | 1.80 | 1.75 |
| $K_{turb}$ | 0.31 | 0.31 | 0.31 |
| $T_b$ | 0.79 | 0.78 | 0.65 |
| $T_c$ | 0.0 | 0.0 | 0.0 |
| $T_{eng}$ | 0.10 | 0.11 | 0.15 |
| $T_{fload}$ | 3.0 | 3.02 | 2.90 |
| $K_{pload}$ | 25.01 | 24.95 | 24.10 |
| $K_{iload}$ | 0.10 | 0.09 | 0.18 |
| $T_{Pelec}$ | 1.10 | 1.12 | 1.01 |
| $L_{dref}$ | 0.90 | 0.91 | 0.83 |
| $r$ | 0.05 | 0.05 | 0.05 |
| $W_{fnl}$ | 0.43 | 0.43 | 0.43 |
| $K_{PA}$ | 3.95 | 3.93 | 3.81 |
| $K_{IA}$ | 2.84 | 2.82 | 2.71 |
| $K_M$ | 1.10 | 1.09 | 1.23 |
| $K_{FF}$ | 1.30 | 1.33 | 1.45 |

## VI. CONCLUSION

IEEE standards for large scale gas turbine-governor and exciter have been studied in this paper to model a high duty gas turbine-governor and excitation system for an installed gas power plant in Iran. Some practical issues about a gas power plant modeling and identification are discussed. An identification procedure based on Cuckoo Search optimization method, as one of the metaheuristic algorithms, has been proposed. GGOV1 and ST6B standard models' parameters have been identified using proposed algorithm. Validation results and error indices show that the proposed identification method successfully estimates standard models' parameters.

As a recommendation, Inlet Guide Van (IGV) model can be included in GGOV1. Adding this part may improve performance of the overall model at lower loads. Actually, IGVs are fully open at the loads beyond the 70% of nominal load and have not any effect at this range.

APPENDIX

Gas turbine and electrical generator main specifications installed in Chabahar power plant are listed below

TABLE IV
GENERATOR AND GAS TURBINE SPECIFICATIONS

| Parameter | Value |
|---|---|
| Prime mover | GT V94.2 |
| Generator Rated power | 200 MVA |
| Turbine Rated Power | 160 MW |
| Rated Voltage | 15.75 KV |
| Rated Current | 7331 A |
| Rated Speed | 3000 rpm |
| Rated Frequency | 50 Hz |
| Excitation System Type | Static |
| Excitation Voltage at Rated Load | 296 V |
| Excitation Current at Rated Load | 1417 A |

ACKNOWLEDGMENT

The primary author wishes to thank all utility members of the Chabahar power plant, and Iran Grid Management Center (IGMC) who supported this research project.